\documentclass[twoside]{article}
\usepackage{fleqn,espcrc2}

\usepackage{graphicx}

\newcommand{\AmS}{{\protect\the\textfont2
  A\kern-.1667em\lower.5ex\hbox{M}\kern-.125emS}}

\title{
Dynamics {\it vs} electronic states of vortex core of high-$T_c$ superconductors investigated by high-frequency impedance measurement
}
\author{A. Maeda\address[DB]{Department of Basic Science, 
    The University of Tokyo, 3-8-1, Komaba, Meguro-ku, Tokyo, 153-8902, Japan}%
        \thanks{Corresponding author.  Tel. +81-3-5454-6747.  Fax. +81-3-5454-6992.  E-{\it mail address}: maeda@maildbs.c.u-tokyo.ac.jp.  Also CREST, Japan Science and Technology Corporation (JST), 4-1-8, Honcho, Kawaguchi, 332-0012, Japan. This work is partially supported by the Grant-in-Aid for Scientific Research on Priority Area (A) No. 258, "Vortex Electronics" sponsored by the Ministry of Education, Science, Sports and Culture.},
        Y. Tsuchiya\addressmark[DB],  K. Iwaya\addressmark[DB], K. Kinoshita\addressmark[DB], T. Hanaguri\address[DAM]{Department of Advanced Material Science, The University of Tokyo, 7-3-1, Hongo, Bunkyo-ku, Tokyo, 113-8586, Japan},
         H. Kitano\addressmark[DB], T. Nishizaki\address[MR]{Institute for Materials Research, Tohoku University, 2-1-1, Katahira, Aoba-ku, Sendai, Miyagi, 980-8577, Japan}, K. Shibata\addressmark[MR], N. Kobayashi\addressmark[MR],
       J. Takeya\address[CRIEPI]{Central Research Institute of Electric Power Industry, 2-11-1, Iwato-kita, Komae, Tokyo, 201-8577, Japan},
        K. Nakamura\addressmark[CRIEPI] and Y. Ando\addressmark[CRIEPI]}

\begin{document}

\begin{abstract}
Dynamics of vortices reflects the electronic states of quasiparticles in the core.  To understand this, we investigated the following three issues.
(1) We investigated the complex surface impedance, $Z_s$, of YBa$_2$Cu$_3$O$_y$ as a function of magnetic field, $H$.
The total features were well expressed by the Coffey-Clem model.  From the data, we estimated the viscosity and pinning frequency, which were found to be independent of frequency.  In particular, the obtained viscosity definitely shows that the core of vortex of YBa$_2$Cu$_3$O$_y$ is moderately clean.
(2) An anomaly found in the surface reactance at the first order transition (FOT) of vortex lattice was investigated in Bi$_2$Sr$_2$CaCu$_2$O$_y$ with various doping levels.  As a result, the anomaly was found only in the samples exhibiting the FOT.  On the other hand, we did not observe the anomaly in YBa$_2$Cu$_3$O$_y$.  These suggest that the anomaly is due to the change in the electronic states of the vortices characteristic of materials with very strong anisotropy.
(3) We measured $H$ dependence of both the thermal conductivity $\kappa(H)$ and $Z_s(H)$ in exactly the same pieces of crystal.  We could not find any anomaly in $Z_s(H)$ even at the onset of the plateau. This result suggests that the origin of the plateau in $\kappa(H)$ is not a drastic phase transition but is rather gradual crossover.

\vspace{1pc}
\end{abstract}

\maketitle

\section{INTRODUCTION}

Vortex motion is closely related to the electronic structure of the vortex core\cite{Golo}.
Since it has been established that the condensate wave function of high-$T_c$ superconductor (HTSC) is mainly $d_{x^2-y^2}$ like, new features are expected in the dynamics of vortices and that of quasiparticles in the mixed state\cite{Wang}.

The second important feature in the mixed state of HTSC is the existence of the first order transition (FOT) in the vortex lattice (VL)\cite{George}.
It is widely accepted that the basic characteristics of the melting transition can be described within the framework of the phenomenological Ginzburg-Landau (GL) theory if the effect of thermal fluctuation is properly taken into account~\cite{Dodgson}.
However, detailed nature of the transition is still to be understood.

Another important aspect of the vortex of HTSC is a quantum nature of the core.  That is, in HTSC, the product $\xi_0k_F$ ($\xi_0$ and $k_F$  are the coherence length and the Fermi wavenumber, respectively) is very close to unity, which is very different from many other superconductors.  
Flux flow was first discussed by Bardeen and Stephen\cite{BS} for dirty $s$-wave superconductors.  Succeedingly, Larkin and Ovchinnikov discussed clean $s$-wave superconductor\cite{LO}.
On the other hand, the flux flow of $d$-wave superdconductor was discussed by Kopnin and Volovik\cite{KV}, who showed that the quasiparticles outside the core were important only for thermodynamic quantity like specific heat as a function of magnetic field, and did not make an important contribution to the flux flow.  Physically, this is because that the dissipative process takes place close to the boundary of the core.
All of these studies were for the classical core with $\xi_0k_F\gg$1.
For quantum core with $\xi_0k_F\sim$1, Larkin and Ovchinikov\cite{LO} discussed the $I-V$ characteristics, and Koulakov and Larkin discussed the microwave response\cite{KL}.   In particular, the latter authors predicted many new resonance structures in the $I-V$ characteristics.
However, these calculations were for $s$-wave superconductors.
No theoretical studies have been made for $d$-wave quantum core.
In the quantum core, the distinction between the quasiparticles in the core and outside the core loses physical meaning.  Thus,  new phenomena are expected for such superconductors.

Based on these backgrounds, we measured the complex surface impedance $Z_s=R_s+iX_s$ of YBa$_2$Cu$_3$O$_y$ (YBCO) single crystals around the vortex melting transition, and estimated the viscosity and pinning frequency of vortices.
In addition, the anomaly found in $X_s$ at the FOT in Bi$_2$Sr$_2$CaCu$_2$O$_y$ (BSCCO) \cite{Hana1} was studied extensively in samples with various doping levels.  We tried to search the same phenomena in YBCO.
We also compared the thermal conductivity data and $Z_s$ data using exactly the same pieces of crystal of BSCCO, and tried to investigate the nature of the anomaly proposed by a recent thermal conductivity measurement in the same material\cite{Kri}.

\section{EXPERIMENTAL}

High-purity untwinned YBCO single crystals were grown by a self-flux method.
It was almost optimally doped, and showed a superconducting transition temperature, $T_c$, of 91 K\cite{Nishi}.
Crystals of BSCCO were fabricated by the FZ method.
Crystals with various doping levels were made by the post annealing in different oxygen atmospheres.  A crystal with the composition Bi$_2$Sr$_2$Ca$_{0.8}$Y$_{0.2}$Cu$_2$O$_y$ (BSCYCO) with $T_c$ of 56 K was also prepared.
In addition to characterizations by dc resistivity- and dc magnetic-susceptibility measurements, the presence of the FOT in the VL was always checked by the magnetization measurement using SQUID magnetometer or local Hall probes\cite{Tsuboi}.

Surface impedance, $Z_s$, was measured by the cavity perturbation method at some microwave and millimeter-wave frequencies.
The surface resistance $R_s$ and the surface reactance $X_s$ can be obtained from the changes in the quality factor of the cavity and the resonance frequency, respectively.
In all the measurements, both static and microwave magnetic fields were applied along the $c$-axis to examine the {\it in-plane} response.

Details for the thermal conductivity measurement were described elsewhere\cite{Ando}.

\section{VISCOSITY AND PINNING FREQUENCY OF YBCO AND BSCCO}

If one knows $Z_s$ as a function of magnetic field, important parameters like viscosity, $\eta$, and pinning frequency, $\omega_p$, can be obtained.
However, there have been many confusions on these values, because of the lack of systematic measurements of the {\it complex} electromagnetic response as a function of magnetic field.
We measured the complex surface impedance of very high-quality untwinned YBCO (almost optimally doped) as a function of magnetic field\cite{Tsuchiya}.
A plot in the impedance plane ($R_s$ {\it vs} $X_s$) can be fitted by a theoretical expectation of Coffey and Clem (CC)\cite{CC} very well.
From the fit, we estimated that $\eta\sim$ 5 $\times$ 10$^{-7}$ Ns/m$^2$, and $\omega_p\sim$ 40 GHz in the low temperature limit.  Both of these parameters decreased with increasing temperature.  It was remarkable that both $\eta$ and $\omega_p$ did not depend on magnetic field nor frequency at least between 19 GHz and 40 GHz.  This shows that the above mentioned procedure and the obtained values are very reliable.
In particular, these values mean that the ratio of the spacing of energy levels $\hbar\omega_0$ and the life time of these levels $\hbar/\tau$, $\omega_0\tau$, is $\sim$ 0.3 in the low temperature limit.
This means that the  vortex core of YBCO is moderately clean.

\begin{figure}[thbp]
\includegraphics[scale=0.38]{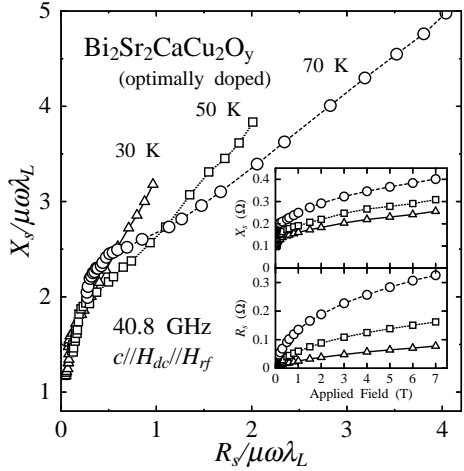}
\caption
{Impedance plane plot of $Z_s$ of a BSCCO crystal with optimum doping.
An inset shows $R_s$ and $X_s$ as a function of dc magnetic field.  Different marks in the inset correspond to those in the main panel.
}
\label{RXBSC}
\end{figure}

\begin{figure}[thbp]
\includegraphics[scale=0.5]{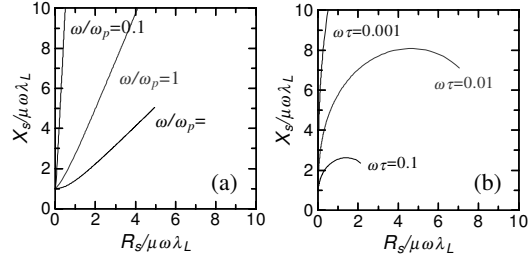}
\caption
{Expected behavior of the surface impedance shown in the impedance plane. (a) vortex dynamics (b) pairbreaking.
}
\label{RXYBCO}
\end{figure}

\begin{figure}[thbp]
\includegraphics[scale=0.37]{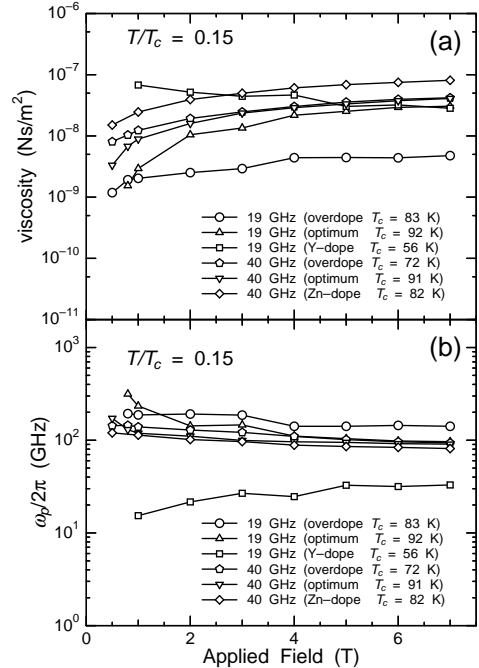}
\caption
{Viscosity and pinning frequency of BSCCO with various doping levels.}
\label{BSCpara}
\end{figure}

We also performed the same measurements in BSCCO with various carrier concentrations.
As is discussed in the next section,
in BSCCO, the increase of the London penetration depth at the FOT\cite{Hana1} makes the problem more complicated.
Because of this phenomenon, the low-field data ($<\sim$ 1 T) shown in Fig.~\ref{RXBSC} did not obeys the expected behavior by the CC model which was shown in Fig.~\ref{RXYBCO}~(a), but rather showed the behavior expected by the pairbreaking scenario (Fig.~\ref{RXYBCO}~(b)).
However, the higher field data ($>\sim$ 1 T) show a tendency to approach the CC behavior asymptotically.
Thus, we believe that the similar estimation makes sense.  In this case, however, $\omega_p$ and $\eta$ become field dependent.
The actual procedure was as follows.  In the CC model, the ratio of the increase of the real part of the resistivity $\Delta\rho_1$ to that of the imaginary part, $\Delta\rho_2$, $\Delta\rho_1/\Delta\rho_2$, is expressed as follows.
\begin{equation}
\frac{\Delta\rho_1}{\Delta\rho_2} = \frac{\omega_p}{\omega},
\label{BSCana}
\end{equation}
where $\omega$ is the angular frequency, and the complex resistivity $\rho$ is related to the complex surface impedance $Z_s$ as
\begin{equation}
Z_s^2 = i\omega\mu_0\rho,
\end{equation}
where $\mu_0$ is the permeability of vacuum.
Using eq.~(\ref{BSCana}), we can determine $\omega_p$ at each magnetic field.
Once $\omega_p$ was fixed, the viscosity $\eta$ could be determined at each field, since both $R_s$ and $X_s$ are known functions of $\omega_p$ and $\eta$.
The resultant parameters are shown in Fig.~\ref{BSCpara}.
As was mentioned above, the parameters were field-dependent.  However, the data above 5 T and 0.5 T did not depend on magnetic field so much, which corresponds to the the fact that the high-field and low-temperature data obey the CC behavior well.
There, 
we obtained that $\eta\sim$ 5$\times$10$^{-8}$ Ns/m$^2$, as an average.
This gives $\omega_0\tau$ $\sim$ 0.1, which also indicates that the vortex core in BSCCO is moderately clean.

It is remarkable that almost all the samples with various different doping levels show similar values for $\omega_0\tau$, which is also similar to that in YBCO.
These suggest that the moderately clean core is quite universal for HTSC.

As was already mentioned in the beginning, the property of $d$-wave quantum core is almost unknown.
In addition, our result clarified that the core is moderately clean. 
Thus, we found that we are in the completely unknown area in terms of the flux flow. 
The flux flow of HTSC should be studied continuously in the future.

\section{ANOMALY IN SURFACE REACTANCE AT THE FOT OF VL}

The basic form of the GL theory is irrespective of the gap symmetry~\cite{Blatter}.
Therefore, the FOT in the VL has not been discussed in relation to the pairing symmetry other than a very few exceptions~\cite{Volovik}. 
However, the vortex melting transition in a $d$-wave superconductor may affect the node excitation, and anomalies in quantities which are sensitive to the electronic states are expected.

Previously\cite{Hana1}, we found that
 $X_s$ of optimally BSCCO showed a clear jump at the field of the FOT of the VL, while no anomalies were found in $R_s$ at the corresponding field.
The jump height in $X_s$, $\Delta X_s$, was almost independent of temperature and $\Delta X_s/X_s=2\sim 4$~\%.
This result strongly suggested that the stepwise change in $X_s$ is related to the vortex melting transition.
Similar but weaker anomaly was found in $X_s(H)$ around the second-magnetization-peak field which was considered as a dimensional crossover field of the vortex system~\cite{Hana1}.
Therefore, we concluded that $X_s$ increased when the 3-dimensional vortex lattice model became no longer valid.
We emphasize here again that the anomalies in $Z_s$ at the vortex phase boundaries appeared only in the reactive response $X_s$.
After discussions, we concluded that the stepwise increase in $X_s$ at the FOT was not originated from the change in the vortex motion but should be ascribed to the increase in the London penetration depth $\lambda_L$.
Since $1/\lambda_L^2$ is proportional to the superfluid density, this result strongly suggested that the superfluid density or the amplitude of the order parameter $|\psi|$ decreased in the vortex liquid phase.
Considering that $\Delta\lambda_L/\lambda_L=\Delta X_s/X_s\sim \Delta (|\psi|^2)/2|\psi|^2$, $\Delta (|\psi|^2)/|\psi|^2$ at the vortex melting transition was estimated to be $4\sim 8$~\%.
In the vortex liquid phase, strong thermal fluctuation may reduce $|\psi|$.
According to the calculation by Herbut and Te{\v s}anovi{\' c}~\cite{Herbut}, $\Delta (|\psi|^2)/|\psi|^2$ at the transition is about 1~\%, which is somewhat smaller than the observed value.
We speculated that the origin of this difference is related to the node excitation in the $d$-wave superconductor.

\begin{figure}[thbp]
\includegraphics[scale=0.38]{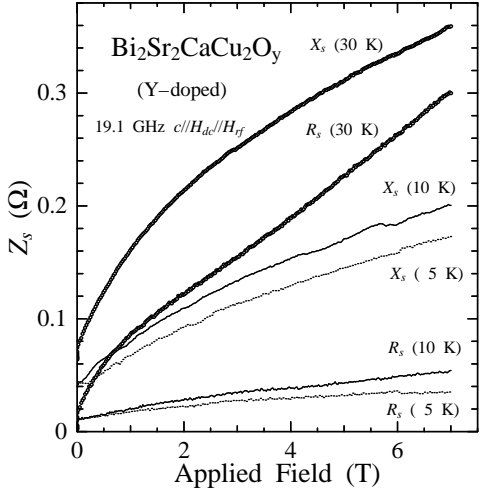}
\caption
{$R_s$ and $X_s$ as a function of magnetic field in Y-substituted BSCCO.}
\label{UDBSC}
\end{figure}

To clarify the origin of the observed anomaly in detail, we investigated the phenomena for wide range of doping level.  We found that the anomaly was found in all samples showing the FOT\cite{HanaMOS}.
Figure~\ref{UDBSC} shows $R_s$ and $X_s$ of an underdoped BSCYCO, which did not show the FOT in the VL.  There, we did not see any anomaly in $X_s(H)$.
Thus, for BSCCO, the anomaly was found to be observed quite universally only in crystals showing the FOT in the VL.

We also investigated the possibility for the similar phenomenon in YBCO\cite{Tsuchiya}.  However, in YBCO, we did not find any anomaly both in $R_s$ and in $X_s$ at the FOT nor at the so-called crossover field\cite{Nishi}.
This suggests that the anomaly found in BSCCO is characteristic of materials with very large anisotropy.
The change in the Josephson coupling between different CuO$_2$ layers at the FOT found in the Josephson plasma resonance experiments\cite{Shiba,Matsuda} might be related to the phenomena discussed here.

\section{THERMAL CONDUCTIVITY {\it vs} SURFACE IMPEDANCE}

Recently, a sharp kink and subsequent plateau structure in the magnetic field dependence of thermal conductivity, $\kappa$, were reported in BSCCO\cite{Kri}, and the occurrence of a new phase transition was suggested in the condensate from the $d_{x^2-y^2}$ state to the fully-gapped state\cite{Kri}.
However, these anomalies were associated with hysteresis\cite{Aubin}.  In addition, in subsequent studies\cite{Ando}, a sharp kink was not reproduced.
Ando {\it et al.}\cite{Ando} also found that the observed plateau feature was quite sensitive to the impurity scattering rate of samples.   Thus, the origin of the anomalies has not been clarified.
If such a sudden change takes place in the electronic state, a similar anomaly should be expected in $Z_s(H)$.

\begin{figure}[thbp]
\includegraphics[scale=1.0]{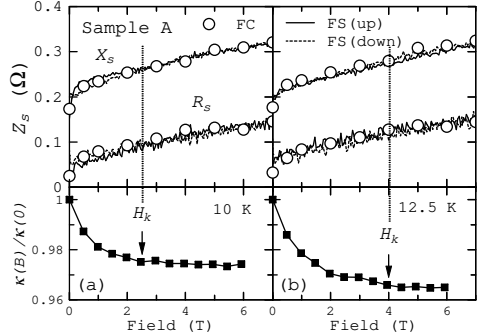}
\caption
{$Z_s(H)$ of a BSCCO sample, together
with the observed plateau structure in $\kappa(H)$. 
Continuous lines and open circles correspond to 
field-swept (FS) data after zero-field cooling and 
 field-cooled (FC) data, respectively. }
\label{kappa1}
\end{figure}


In our samples, we never observed any sharp kink structure in the field dependence of $\kappa$.  However, 
in some sample, the plateau structure was found.
Figure~\ref{kappa1} shows the magnetic-field dependence of $Z_s$ at two different temperatures, together with the magnetic-field dependence of thermal conductivity $\kappa(H)$.  The plateau structure was observed in $\kappa(H)$.
Except for the low-field region below 0.5 T, both the field-swept (FS) and field cooled (FC) data coincided with each other, which suggested that the field inhomogeneity did not play a serious role in the high-field region. 
Therefore, we discuss the $Z_s(H)$ profile based on the FC data.

As was already discussed, in general, $Z_s$ of the type-II superconductor is described by several possible vortex dynamics parameters, such as $\omega_p$, $\eta$, creep factor, quasi-particle density, and $\lambda_L$\cite{CC}. 
 Since the vortex motion was closely related to the electronic structure of vortex core, we expect an anomaly in the microwave response if a drastic change takes place in the electronic structure. 
At both temperatures, however, both $R_s$ and $X_s$ increased monotonically with increasing field, and at $H_k$'s, defined as the field above which the plateau started in $\kappa(H)$ at each temperature, no distinct anomaly was observed in $Z_s(H)$.

It is also noted that another sample which did not show the ``plateau behavior" exhibited a very similar $Z_s(H)$ to that of the sample discussed just above.
These results  imply that the phase transition is unlikely to be an origin of ``the plateau" and "the possible sharp kink" in $\kappa(H)$.

\section{CONCLUSION}

To understand the electronic structure of the vortex of HTSC and its influence on the dynamics, we investigated the high-frequency surface impedance $Z_s$, and found the followings.
(1) $Z_s(H,T)$ of YBCO  was well expressed by the CC model.  From the data, we estimated $\omega_p$ and $\eta$, which were found to be independent of magnetic field and of frequency investigated.  In particular, the obtained viscosity, $\eta$, definitely showed that the core of vortex of YBCO was moderately clean.
We also found that the vortex core of BSCCO was moderately clean.
(2) An anomaly found in $X_s(H)$ at the FOT of VL was investigated in BSCCO with various doping levels.  We found that the anomaly was found only in the crystals which exhibit the FOT in the VL.  On the other hand, we did not observe the anomaly in YBCO.  These suggest that the anomaly is due to the change in the electronic states of the vortices, characteristic of materials with very strong anisotropy.
(3) We measured the $H$ dependence of both the thermal conductivity $\kappa(H)$ and microwave surface impedance $Z_s(H)$ in exactly the same peaces of crystal. In all samples, we could not find any anomaly in $Z_s(H)$ at the onset of ``the plateau". This result suggests that the origin of the plateau and the kink reported previously in $\kappa(H)$ is not a phase transition.


\begin{thebibliography}{99}

\bibitem{Golo} See for instance: M. Golosovsky, M. Tsindlekht, and D. Davidov, Superconud. Sci. Technol., 9 (1996) 1.

\bibitem{Wang} Y. Wang amd A. H. MacDonald, Phys. Rev. B52 (1995) R3876.


\bibitem{George} {\it See for instance}, G. W. Crabtree and D. R. Nelson, Physics Today, 1997 April issue, pp.38.

\bibitem{Dodgson} Matthew J. W. Dodgson {\it et al.}, Phys. Rev. Lett., 80 (1998) 837.

\bibitem{BS} J. Bardeen and M. J. Stephen, Phts. Rev. 140 (1965) A1197.

\bibitem{LO} A. Larkin and Yu. Ovchinnikov, Pis'ma Zh. Eksp. Teor. Fiz. 23 (1976) 210 [Sov. Phys. JETP Lett. 23 (1976) 187].

\bibitem{KV} N. B. Kopnin and A. V. Volovik, Phys. Rev. Lett. 79 (1997) 1377.

\bibitem{KL} A. A. Koulakov and A. I. Larkin, Phys. Rev. B59 (1999) 12021.

\bibitem{Hana1} T. Hanaguri {\it et al.}, Phys. Rev. Lett., 82 (1999) 1273.

\bibitem{Nishi} T. Nishizaki {\it et al.}, Phys. Rev. B58 (1998) 11169.

\bibitem{Kri} K. Krishana {\it et al.}, Science, 277 (1997) 83.

\bibitem{Tsuboi} T. Tsuboi, T. Hanaguri, and A. Maeda, Phys.  Rev. Lett., 80 (1998) 4550.

\bibitem{Ando} Y. Ando {\it et al.}, Phys. Rev. B62 (2000) 626.

\bibitem{Tsuchiya} Y. Tsuchiya {\it et al.}, Advances in Physics XII (Springer, 2000) p. 66-69, also a paper in this volume.

\bibitem{CC} M. W. Coffey and J. R. Clem, Phys. Rev. Lett., 67 (1991) 386.

\bibitem{Blatter} G. Blatter {\it et al.}, Rev. Mod. Phys., 66 (1994) 1125.

\bibitem{Volovik} G. E. Volovik, {\it JETP Lett.},  65 (1997) 491.

\bibitem{Herbut} Igor F. Herbut and Zlatko Te{\v s}anovi{\' c}, Phys. Rev. Lett.,  73 (1994) 484.

\bibitem{HanaMOS} T. Hanaguri {\it et al.}, J. Low Temp. Phys. 117 (1999) 1405.

\bibitem{Shiba} M. Sato {\it et al.}, Phys. Rev. Lett. 79 (1997) 3759.

\bibitem{Matsuda} Y. Matsuda {\it et al.}, Phys. Rev. Lett. 75 (1995) 4512.

\bibitem{Aubin} H. Aubin {\it et al.}, Phys. Rev. Lett. 82 (1999) 624.


\end{thebibliography}
\end{document}